\documentclass[
10pt,journal
]{IEEEtran}
\usepackage[latin1]{inputenc}
\usepackage[boxed]{algorithm2e}
\usepackage{amsmath}\usepackage{amsthm}\usepackage{amssymb}\usepackage{mathrsfs}\usepackage{amscd}\usepackage{graphics}\usepackage{graphicx}\usepackage[tight]{subfigure}\usepackage{amsfonts}\usepackage{amsxtra}\usepackage{color}
\usepackage{multirow}
\usepackage{amscd}

\DeclareMathOperator{\R}{\mathbb{R}}

\DeclareMathOperator{\Null}{null}

\DeclareMathOperator{\trace}{trace}

\DeclareMathOperator{\diag}{diag}

\theoremstyle{definition}\newtheorem{definition}{Definition}[section]\newtheorem{remark}[definition]{Remark}
\theoremstyle{plain}\newtheorem{theorem}[definition]{Theorem}\newtheorem{corollary}[definition]{Corollary}

\ifCLASSOPTIONcompsoc
\else
\fi

\ifCLASSINFOpdf
\else
\fi

\begin{document}
%
\title{Schroedinger Eigenmaps for the Analysis of Bio-Medical Data}
%
%
%
%

\author{Wojciech Czaja \IEEEcompsocitemizethanks{\IEEEcompsocthanksitem W.~Czaja is with University of Maryland, Department of Mathematics, College Park, MD 20742, wojtek@math.umd.edu.} and Martin Ehler \IEEEcompsocitemizethanks{\IEEEcompsocthanksitem M.~Ehler is with Helmholtz Zentrum M\"unchen, German Research Center for Environmental Health, Institute of Biomathematics and Biometry, Ingolst\"adter Landstr.~1, 
D-85764 Neuherberg, martin.ehler@helmholtz-muenchen.de. He is also with National Institutes of Health, Eunice Kennedy Shriver National Institute of Child Health and Human Development, Section on Medical Biophysics, 9 Memorial Drive, Bethesda, MD 20892, ehlermar@mail.nih.gov.
}
}

\markboth{
}
{Shell \MakeLowercase{\textit{et al.}}: Bare Demo of IEEEtran.cls for Computer Society Journals }

\IEEEcompsoctitleabstractindextext{%
\begin{abstract}

We introduce Schroedinger Eigenmaps, a new semi-supervised manifold learning and recovery technique. This method is based on an implementation of graph Schroedinger operators with appropriately constructed barrier potentials as carriers of labeled information. We use our approach for the analysis of standard bio-medical datasets and new multispectral retinal images.

\end{abstract}

}

\maketitle

\IEEEdisplaynotcompsoctitleabstractindextext

%
\IEEEpeerreviewmaketitle


\section{Introduction}

Typical modern bio-medical data are characterized by high complexity (usually by means of high dimensionality), and the underlying nonlinear processes are often unknown, forming a fundamental barrier to better understand human physiology. Moreover, as personalized medicine expands, increasingly detailed patient data need to be integrated in clinical trials and individual treatment decisions, which usually results in a complicated classification problem.  

Data driven analysis schemes are commonly based on kernels that derive the proper classification directly from the data. However, bio-medical applications are also characterized by low signal to noise ratio so that effective schemes require additional expert input. This outside information is, depending on the perspective, called labeling, (semi-) supervision, regularization, or learning from training data. Often, because of the nature of medical experiments, there are only few training points available and analysis schemes are needed that learn from these few data points most efficiently, while being robust against perturbations.

%

%

Examples of kernel based analysis techniques are Locally Linear Embedding (LLE) \cite{Roweis:2000aa}, Hessian-based LLE \cite{Donoho:2003ab}, Laplacian Eigenmaps \cite{Belkin:2003aa}, Diffusion Wavelets \cite{Coifman:2005aa,Coifman:2006ab}, or Diffusion Maps \cite{Coifman:2006ac}, and kernel PCA and Support Vector Machines \cite{Scholkopf:2002fk}. These methods represent the data in form of a graph, with nodes that are formed by the data vectors, and with edges that represent the distances between pairs of such vectors. This information is stored in the adjacency matrix, which then is modified to form the kernel. 


Physical or experimental constraints often suggest that the data points belong to a low-dimensional manifold. Many kernel-based methods recover this manifold by means of representations in terms of the most significant eigenvectors of the kernel matrix \cite{Tenenbaum:2000aa}. A major feature of Laplacian Eigenmaps and related methods is the preservation of local spectral distances, while reducing the overall dimension of the data represented by the aforementioned graph, cf., \cite{Chung:1997aa}. It has been successfully applied to pattern recognition and segmentation problems \cite{Sundaresan:2008aa,Wilson:2005aa} and \cite{Bachmann:2005aa,Benedetto:2009ac}.

Laplacian Eigenmaps and similar methods originally lead to fully automated classification algorithms but have also been modified to work in semi-supervised settings:~Belkin and Niyogi \cite{Belkin:2002ys}, \cite{Belkin:2004vn}, for instance,  proposed to use classifiers induced from partially labelled data on the Laplacian representation of unlabeled points. Belkin et al.~\cite{Belkin:2006aa} used regularization in reproducing kernel Hilbert spaces to obtain a family  of semi-supervised and transductive learning algorithms related to Laplacian Eigenmaps. In \cite{Kim:2009zr}, extensions of the regularization methods proposed in \cite{Belkin:2006aa} are obtained via inverse regression. In related developments, Zhu et al.~\cite{Zhu:2003ys} proposed to augment the data graph with the results of a prior classifier. Local discriminant analysis to locally modify (``warp") the diffusion metric is proposed by Coifman et al.~in \cite{Coifman:2005aa}. Szlam et al.~\cite{Szlam:2008kx} developed a further generalization 
by means of collecting all the class information into a modification of the metric used to build the kernel.

In the present paper we develop a new \emph{Schroedinger Eigenmaps algorithm} that builds upon Laplacian Eigenmaps and steers the diffusion process within the uniform theory of Schroedinger operators in place of the Laplacian. Experts (e.g., trained practitioners) can introduce their input in form of a barrier potential for the associated Schroedinger operator on a graph to improve the detection and classification processes by steering the Schroedinger diffusion process effectively. 
%
At the same time, the new graph Schroedinger operator converges towards the Schroedinger operator on the manifold (just like the Laplacian kernel converges towards the Laplace-Beltrami operator), see Section \ref{subsection schroedinger}. 

Such defined barrier Schroedinger potentials are easy to implement and appear in our numerical experiments to be well-suited for the analysis of low quality data with only few training points available -- a typical situation in bio-medical applications. After having validated the usefulness of Schroedinger Eigenmaps (SE) by comparison to Support Vector Machines (SVM) on three standard medical datasets, we apply our method to analyze multispectral retinal images and genetic expression profiles. The proposed scheme provides two major features for data segmentation. On one hand, by an appropriate selection of the potential, it allows the practitioner to separate data vectors, which should not appear in the same class. Specifically, in bio-medical imaging, the practitioner could be a physician who has additional information from results of other imaging devices or from physiological processes and knowledge about disease progression. 
On the other hand, Schroedinger Eigenmaps allow to identify complex regions by means of decreasing the relative distances between labeled representatives. For the physician who has already identified several different regions of pathology, this technique allows to collect these regions into one pathology class and therefore allows to better indentify normal physiology.

In Section \ref{section laplacian} we review Laplacian Eigenmaps. We introduce the new Schroedinger Eigenmaps with an investigation of its properties in Section \ref{section schroedinger type}, where we also visualize the behavior on artificial data. Section \ref{section classification} is dedicated to the description of the intertwining of Schroedinger Eigenmaps and classification schemes. In Section \ref{section experiments}, we compare Schroedinger Eigenmaps with Support Vector Machines on $3$ standard medical datasets. We also analyze new multispectral retinal images.

%
%



\section{Laplacian Eigenmaps}\label{section laplacian}
Given $m$ points $\{x_1,\ldots,x_m\}\subset\mathbb{R}^N$, we assume that they belong to an $n$-dimensional manifold, where $n$ is much smaller than $N$. The goal is to recover this manifold or, in other words, to find a low-dimensional representation $\{y_1,\ldots,y_m\}\subset\mathbb{R}^n$ of the original dataset $\{x_1,\ldots,x_m\}\subset\mathbb{R}^N$. Here, we briefly recall the three-step procedure of Laplacian Eigenmaps, see \cite{Belkin:2003aa}: 

\textbf{Step 1.} \textit{Adjacency graph construction
:}\\
Build a graph $\mathcal{G}$, whose nodes $i$ and $j$ are connected if $x_i$ is among the $k$-nearest neighbors of $x_j$ or vice versa. The distances between data points are measured in the Euclidean metric. The graph $\mathcal{G}$ represents the connectivity of the data, and $k$ is a parameter.

\textbf{Step 2.} \textit{Heat kernel as weights:}\\
Weight the edges of the graph. A typical example is the \emph{diffusion weight matrix} $W$ given by
\begin{equation*}
W_{i,j}=\begin{cases}
e^{-\frac{\|x_i-x_j\|^2}{\sigma}} ,& \text{$i$ and $j$ are connected},\\
0, & \text{otherwise.}
\end{cases}
\end{equation*}
\textbf{Step 3.} \textit{Solving an eigenvalue problem:}\\
Let $D$ be a diagonal matrix with $D_{i,i}=\sum_j W_{i,j}$. Denote the low-dimensional representation by an $m \times n$ matrix $y=(y_1,\ldots,y_m)^\top$, where each row $y_i$ is considered as a vector in $\mathbb{R}^n$. Next, consider the following minimization problem
\begin{equation}\label{laplace mini 1}
\min_{y^\top Dy=I}\tfrac{1}{2}\sum_{i,j}\|y_i-y_j\|^2 W_{i,j}=\min_{y^\top Dy=I} \!\trace(y^\top L y),
\end{equation}
where $L=D-W$ is the $m \times m$ Laplace operator and $I$ is the identity matrix. The minimizer of \eqref{laplace mini 1} is given by the $n$ minimal eigenvalue solutions of $Lx=\lambda Dx$ under the constraint $y^\top D y=I$, i.e., the columns of the minimizer $y^{(0)}$ are the $n$ eigenvectors with respect to the smallest eigenvalues. If the graph is connected, then ${\bf 1}=(1,\ldots,1)^\top$ is the only eigenvector with eigenvalue $0$. Instead of \eqref{laplace mini 1}, we solve for $\min\trace( y^\top L y)$ subject to $y^\top Dy=I$, $y^\top D^{1/2}{\bf 1}=0$. 
By applying $z=D^{1/2}y$, this yields $\min\trace(z^\top \mathcal{L} z)$ subject to $z^\top z=I$, $z^\top {\bf 1}=0$, 
where $\mathcal{L}=D^{-1/2} L D^{-1/2}$. The minimizer $z^{(0)}$ is given by the $n$ eigenvectors with smallest nonzero eigenvalue, and we obtain the sought $n$-dimensional representation $\{y_1,\ldots,y_m\}$ from $y=D^{-1/2}z^{(0)}$.

Though algorithms based on spectral decomposition, such as Laplacian Eigenmaps, provide
a powerful tool for non-linear dimensionality reduction and manifold learning, they are not without certain shortcomings. We refer the interested reader to, e.g., \cite{Gerber:2007uq},  \cite{Goldberg:2008uq}, \cite{Trosset:2010fk}. 



\section{Schroe\-dinger Eigenmaps}\label{section schroedinger type}
The idea of using solutions of the Laplace equation for labeling is not new. In \cite{Grady:2006aa}, for instance,  
the concept is utilized for image segmentation, where a small number of pixels is labeled, and random walk theory is then applied to classify unlabeled surrounding pixels. We aim to develop a more general classification framework by means of Schroedinger operators. This approach allows for more flexible labeling schemes. Moreover, a significant advantage of the Schroedinger Eigenmaps technique is that it applies not just to surrounding pixels, but to any image location allowing for classes whose members are not spatially connected. This more general approach enables its application to data that do not encode location as, for instance, gene expressions.

The motivation behind this approach is two-fold. Mathematically, we are interested in shifting the analysis of  complex, high-dimensional phenomena towards studying the associated operators acting on these data-dependent graphs, and away from analyzing pointwise relationships between the graph nodes. Physically, partially labeled sets naturally lead to the notion of barrier potentials and to the way in which these potentials affect the diffusion processes on graphs. As such, by appropriately choosing locations of the potential barriers we can steer the diffusion process to allow for identification of the correct cluster containing the labels, when compared to unobstructed diffusion process of the Laplacian Eigenmaps algorithm.

We would like to point out that \cite{Aubry:2011kx} and \cite{Bronstein:2011uq} deal with Schroedinger equations with Hamiltonians, which are Laplacians without any potential. We, on the other hand, deal with Schroedinger operators,  which are Laplacians with barrier potentials; as such, these are two different generalizations of the Laplace operator approach.


\subsection{From Laplace to Schroedinger}
The matrices $L$ and $\mathcal{L}$ in the previous section can be associated with the notion of the Laplace operator $\Delta$, cf.~\cite{Belkin:2003aa}. 
The related diffusion equation 
$ 
\partial_t \varphi = \Delta \varphi 
$ 
is extended to the Schroedinger Equation 
\begin{equation}\label{eq:schroedinger}
\partial_t\psi(x,t) =  \Delta \psi(x,t) + v(x)\psi(x,t)
\end{equation}
by adding a potential term $v(x)$ to the Laplace operator.  
We revisit this idea in the following section to introduce a discrete Schroedinger operator.


\subsection{Schroedinger Eigenmaps}\label{subsection schroedinger}
The potential $v$ in \eqref{eq:schroedinger} is considered as a nonnegative multiplier operator. The discrete analogue of $\mathcal{E}= \Delta+v$ is the matrix $E=L+V$, where $V$ is a nonnegative diagonal $m \times m$ potential matrix. We replace \eqref{laplace mini 1} with
\begin{equation}\label{schroedinger mini 1}
\min_{y^\top Dy=I}\trace(y^\top (L+\alpha V) y),
\end{equation}
which is equivalent to 
\begin{equation*}
\min_{y^\top Dy=I}[\trace(y^\top L y) + \trace(y^\top \alpha V y)].
\end{equation*}
The parameter $\alpha\geq 0$ is added here so that it can be used to emphasize the relative significance of the potential $V$ with respect to the graph Laplace operator. 

According to \eqref{laplace mini 1}, the minimization problem \eqref{schroedinger mini 1} is equivalent to
\begin{equation}\label{schroedinger mini 10}
\min_{y^\top Dy=I}\tfrac{1}{2}\sum_{i,j}\|y_i-y_j\|^2 W_{i,j} + \alpha \sum_i V(i) \|y_i\|^2,
\end{equation}
where $V=\diag(V(1),\ldots,V(m))$. 
The first component of the above sum incurs a penalty, when neighboring points $x_i$ and $x_j$ are mapped into points $y_i$ and $y_j$, respectively, which are far apart. The second component penalizes these points $y_i$, $i=1, \ldots, m$, which are associated with large values of $V(i)$.
Alternatively speaking, if $V$ took only two values, $0$ and $1$, then the minimization \eqref{schroedinger mini 10} yields a dimension-reduced representation $y$, which forces  increased clustering of the representations $y_i$ of points associated with the value $V(i)=1$, while attempting to ensure that close points remain close after the dimension reduction. 
As such we may utilize the potential $V$ to \emph{label} points which we want to be identified together after the dimension reduction. Because of the built-in preservation of topology of the point cloud (induced by the Laplacian), this labeling may be used to segment a particular class of points.
We shall make this intuition clear in the following sections, see, e.g., Corollary \ref{corollary:direct} and Corollary \ref{corollary:indirect}, or the applications in Section \ref{section experiments}.

It is known that the (rescaled) graph Laplacian converges towards the Laplace-Beltrami operator on the underlying manifold, e.g., \cite{Belkin:2008aa,Luxburg:2008fk}. This convergence carries over to the Schroe\-dinger operator as we shall see next. The Laplacian map is given by 
\begin{equation*}
L^\sigma_m f (x_i)=\!f(x_i)\sum_j e^{-\frac{\|x_i-x_j\|^2}{\sigma}}- \!\!\sum_j f(x_j)e^{-\frac{\|x_i-x_j\|^2}{\sigma}}\!\!.
\end{equation*}
For the purpose of this analysis we choose $k=m$, i.e., we assume that all data points are connected. Replacing the $x_i$ by an arbitrary $x\in\R^n$ yields the extension to $\R^d$:
\begin{equation*}
L^\sigma_m f (x)=f(x)\sum_j e^{-\frac{\|x-x_j\|^2}{\sigma}}-\sum_j f(x_j)e^{-\frac{\|x-x_j\|^2}{\sigma}}.
\end{equation*}
It turns out that there exists a positive constant $C$ such that for i.i.d.~uniformly sampled data points $\{x_1,\ldots,x_m\}$ and $\sigma_m=4m^{-\frac{1}{n+2+s}}$, where $s>0$, and $f\in\mathcal{C}^\infty(\mathcal{M})$, we have the convergence 
\begin{equation}\label{convergence 1}
\lim_{m\rightarrow\infty} C\frac{(\pi \sigma_m)^{-\frac{n+2}{2}}}{m}L^{\sigma_m}_m f (x)=\Delta_{\mathcal{M}}f(x)
\end{equation}
in probability, cf.~\cite{Belkin:2008aa}. Let $v$ be a given potential on the manifold $\mathcal{M}$. The associated matrix $V$ acting on a discrete $m$-point cloud is defined as $V=\diag(v(x_1),\ldots,v(x_m))$. 
Since the potential does not depend on $\sigma$,
we may replace $x_i$ by an arbitrary $x\in\R^n$ to obtain the map:
\begin{equation*}
V_m f(x)=v(x)f(x).
\end{equation*}
Clearly this extension coincides with the continuous potential on the manifold. As such, adding the discrete potential $V_m$ to the discrete Laplacian does not impede the convergence in \eqref{convergence 1}. Consequently, the term
\begin{equation}\label{convergence Schroedinger}
C\frac{(\pi \sigma_m)^{-\frac{n+2}{2}}}{m}L^{\sigma_m}_m f (x)+V_m f(x)
\end{equation}
converges for $n\rightarrow \infty$ towards
\begin{equation*}
\Delta_{\mathcal{M}}f(x) + v(x)f(x)
\end{equation*}
in probability. 
For more results on convergence of the Schroedinger Eigenmaps algorithm to Schroedinger operators on manifolds, we refer the interested reader to \cite{Halevy:2011kx} and \cite{Czaja:2011fk}.

We close this section by noting that expression \eqref{convergence Schroedinger} induces a specific choice of the parameter $\alpha$. Indeed, in order to consider $E=L+\alpha V$, rescaling of $L^{\sigma_m}_m$ in \eqref{convergence Schroedinger} implies the need to reversely rescale $V$. This can be done by means of multiplication with $\alpha= \frac{1}{C}\frac{m}{(\pi \sigma_m)^{-\frac{n+2}{2}}}$, which tends to infinity as $m\rightarrow \infty$. As such, optimal selection of $\alpha$ is related to the size of the dataset.


\subsection{General Properties}\label{sec:general}
In this section we study the properties of the Schroedinger Eigenmaps (SE). We shall see that the potential can be used to push vectors towards zero, which can be helpful in classification tasks. In fact, we shall see later that even nondiagonal ``potentials'' make sense and, thus, address a more general setting where $E=L+\alpha V$ is symmetric positive semi-definite. Note that $E$ can still be diagonalized and all eigenvalues are nonnegative. If $V\neq 0$ is a nonnegative diagonal potential, then ${\bf 1}\not\in\Null(V)$. For a connected graph $\mathcal{G}$, the latter implies that $E=L+\alpha V$ has full rank, i.e., there is no eigenvalue that equals zero.  

To study the effect of the potential $V$ on eigenvectors of the Schroedinger operator, we make use of the notation $\|y\|^2_M:=\trace(y^\top M y)$ for a matrix $M$ and we denote $\mathcal{V}:= D^{-1/2} V D^{-1/2}$:
\begin{theorem}\label{theorem:global potential estimates}
Let $V$ be symmetric positive semi-definite, $n\leq \dim(\Null(V))$, and let $y^{(\alpha)}$ be a minimizer of \eqref{schroedinger mini 1}. 
\begin{itemize}
\item[\textnormal{a)}] There is a constant $C_1\geq 0$ such that 
\begin{equation}\label{theorem eq 1}
\|y^{(\alpha)}\|^2_V \leq C_1\frac{1}{\alpha}.
\end{equation}
\item[\textnormal{b)}] If $z^{(\alpha)}_{\Null}$ is the columnwise orthogonal projection of $z^{(\alpha)}=D^{1/2}y^{(\alpha)}$ onto $\Null(\mathcal{V})\neq \{0\}$, then there is a constant $C_2\geq 0$ such that 
\begin{equation}\label{theorem eq 2}
\|z^{(\alpha)}_{\Null}\|_I^2\geq n-C_2\frac{1}{\alpha}.
\end{equation}
\end{itemize}
\end{theorem}

\begin{IEEEproof}
First, we address part (a). 
Applying $z=D^{1/2}y$ to \eqref{schroedinger mini 1} yields
\begin{equation}\label{schroedinger mini 3}
 \min_{z^\top z=1}\trace(z^\top (\mathcal{L}+\alpha\mathcal{V}) z).
\end{equation}
Since $\mathcal{V}$ is symmetric, \eqref{schroedinger mini 3} can be rewritten as
\begin{equation*}
\min_{\substack{z_{\Null}\in\Null(\mathcal{V}),\; z^\perp_{\Null}\in\Null(\mathcal{V})^\perp \\ z=z_{\Null}+z_{\Null^\perp},\; z^\top z=I }}  \trace(z^\top \mathcal{L} z + z^\top_{\Null^\perp} \alpha\mathcal{V}z_{\Null^\perp}),
\end{equation*}
where $z_{\Null}$ is a matrix whose columns are contained in $\Null(\mathcal{V})$. The constant 
\begin{equation*}
C_1:=\min_{\substack{z^\top_{\Null}z_{\Null}=I \\ z_{\Null}\in\Null(\mathcal{V})}} \trace(z_{\Null}^\top \mathcal{L} z_{\Null})
\end{equation*}
only depends on the nullspace of $\mathcal{V}$, and it is associated with the choice $z_{\Null^\perp}=0$. Since $y^{(\alpha)}$ is a minimizer of \eqref{schroedinger mini 1}, $z^{(\alpha)}=D^{1/2}y^{(\alpha)}$ is a minimizer of \eqref{schroedinger mini 3}. Obviously, we have $\|z^{(\alpha)}\|^2_\mathcal{L}+\|z^{(\alpha)}\|^2_{\alpha\mathcal{V}}\leq C_1$, and according to $\|z^{(\alpha)}\|^2_{\mathcal{L}}\geq 0$, we obtain $\|z^{(\alpha)}\|^2_{\alpha\mathcal{V}}\leq C_1$. This finally yields $\|z^{(\alpha)}\|^2_{\mathcal{V}}\leq C_1\frac{1}{\alpha}$. Since $\|z^{(\alpha)}\|^2_{\mathcal{V}}=\|y^{(\alpha)}\|^2_{V}$, we have proven \eqref{theorem eq 1}.

In order to verify \eqref{theorem eq 2}, we will use  
\begin{equation*}
{z^{(\alpha)}}^\top_{\Null}z^{(\alpha)}_{\Null}=I-{z^{(\alpha)}}^\top_{\Null^\perp}z^{(\alpha)}_{\Null^\perp}.
\end{equation*}
Note that  $\mathcal{V}$ is symmetric positive semi-definite and therefore $\|\cdot\|_\mathcal{V}$ is a norm on $\Null(\mathcal{V})^\perp$. Since all norms on a finite dimensional vector space are equivalent, there is a constant $C>0$ such that $\|\cdot\|_I \leq C\|\cdot \|_{\mathcal{V}}$ on $\Null(\mathcal{V})^\perp$.  By applying $C_2:=C^2C_1$, we get the estimate
\begin{align*}
\|z^{(\alpha)}_{\Null}\|_I^2&=n-\|z^{(\alpha)}_{\Null^\perp}\|_I^2 \geq n-C^2\|z^{(\alpha)}_{\Null^\perp}\|^2_{\mathcal{V}}\\
&=n-C^2\|z^{(\alpha)}\|^2_{\mathcal{V}}
=n-C^2\|y^{(\alpha)}\|^2_V \\
& \geq n-C_2\frac{1}{\alpha}.\qedhere
\end{align*}

\end{IEEEproof}
Let us now state the consequences of Theorem \ref{theorem:global potential estimates} for a diagonal potential $V$. 
\begin{corollary}\label{corollary:direct}
Under the assumptions and notation of Theorem \textnormal{\ref{theorem:global potential estimates}}, let $V=\diag(v_1,\ldots,v_m)$, then we have, for all $i=1,\ldots,m$, 
\begin{equation*}
v_i \|y^{(\alpha)}_i\|^2\leq \sum_{i=1}^m v_i \|y^{(\alpha)}_i\|^2=\| y^{(\alpha)}\|_V^2\leq C_1 \frac{1}{\alpha}.
\end{equation*}
\end{corollary}

\begin{remark}\label{remark:counter example}
Let $V=\diag(1,0,\ldots,0)$, then $y^{(\alpha)}_1$ is pushed to zero according to Corollary \ref{corollary:direct}. If $x_1$ and $x_2$ were sufficiently close, then we expect that $y^{(\alpha)}_2$ is also pushed to zero due to the linkage induced by $L$.

\end{remark}

Corollary \ref{corollary:direct} means that a diagonal potential pushes entries of $y^{(\alpha)}$ to zero. On the other hand, the potential does not force any collapse:
\begin{corollary}\label{corollary:indirect}
Under the assumptions and notation of Theorem \textnormal{\ref{theorem:global potential estimates}}, let $V=\diag(v_1,\ldots,v_m)$. 
\begin{itemize}
\item[\textnormal{a)}] If $v_1,\ldots,v_r\neq 0$ and $v_{r+1},\ldots,v_m=0$, then
\begin{equation*}
\sum_{i=r+1}^m D_{i,i}\|y^{(\alpha)}_i\|^2\geq n-\max_{i=1,\ldots,r}\{ \tfrac{D_{i,i}}{v_i}\} C_1 \frac{1}{\alpha}.
\end{equation*}
\item[\textnormal{b)}]  If $v_1=D_{1,1},\ldots,v_r=D_{r,r}$ and $v_{r+1},\ldots,v_m=0$, then
\begin{equation*}
\sum_{i=r+1}^m v_i\|y^{(\alpha)}_i\|^2\geq n-C_1 \frac{1}{\alpha}.
\end{equation*}
\end{itemize}
\end{corollary}
\begin{IEEEproof}
Part (b) is a special case of part (a). To prove part (a), we apply Corollary \ref{corollary:direct}, which yields 
\begin{align*}
\sum_{i=r+1}^m &D_{i,i}\|y_i^{(\alpha)}\|^2  = \sum_{i=r+1}^m\|z^{(\alpha)}_i\|^2=\|z^{(\alpha)}_{\Null}\|_I^2 \\
& = n-\|z^{(\alpha)}_{\Null^\perp}\|_I^2 
= n-\sum_{i=1}^r D_{i,i}\|y^{(\alpha)}_i\|^2 \\
&\geq n-\sum_{i=1}^r \frac{D_{i,i}}{v_i} v_i \|y^{(\alpha)}_i\|^2 \\
& \geq n-\max_{i=1,\ldots,r}\{ \tfrac{D_{i,i}}{v_i}\} \sum_{i=1}^r v_i \|y^{(\alpha)}_i\|^2  \\
&\geq n-\max_{i=1,\ldots,r}\{ \tfrac{D_{i,i}}{v_i}\}C_1\frac{1}{\alpha}.
\end{align*}

\end{IEEEproof}
Corrollary \ref{corollary:direct} says that penalized locations are pushed to zero, and Corollary \ref{corollary:indirect} ensures that not all unlabeled data are collapsed into zero. Thus, labeling can lead to a separation between labeled and certain unlabeled data.




\subsection{Schroedinger Eigenmaps: Nondiagonal Potentials}\label{sec:nondiag}
While a diagonal potential $V$ is used to overweight or underweight certain nodes $y_i$, it turns out that nondiagonal ``potentials'' are useful as well. For $i\neq j$, let $V^{(i,j)}$ be the matrix with entries $V^{(i,j)}_{i,i}=V^{(i,j)}_{j,j}=1$ and $V^{(i,j)}_{i,j}=V^{(i,j)}_{j,i}=-1$ and zeros elsewhere. 
One then verifies that 
\begin{equation*}
\trace(y^\top \alpha V y)= \alpha \|y_i-y_j\|^2,
\end{equation*}
and Theorem \ref{theorem:global potential estimates} yields
\begin{equation*}
\|y^{(\alpha)}_i-y^{(\alpha)}_j\|^2\leq C_1\frac{1}{\alpha}.
\end{equation*}
In other words, this kind of potential allows to identify points. 
Adding the matrix $V^{(i,j)}$ to the Laplacian has the flavor of modifying the underlying graph $\mathcal{G}$ by reweighting its edges. It allows for modifying the weights $W_{i,j}$, $i\neq j$, and we are able to penalize distances between data points which are not neighbors. We note that a similar change in the graph structure results from warping or weighted diffusion as proposed e.g., in \cite{Coifman:2005aa} or \cite{Szlam:2008kx}.

\begin{remark}
If the graph $\mathcal{G}$ is connected, then $E=L+\alpha V^{(i,j)}$ has a simple eigenvalue equals zero since ${\bf 1}\in\Null(V^{(i,j)})$. Thus, we should use the constraint $y^\top D^{1/2}{\bf 1}=0$ for SE with potentials $V^{(i,j)}$. To establish a uniform methodology of Schroedinger Eigenmaps, we use the constraint $y^\top D^{1/2} w=0$ throughout, where $w$ is an eigenvector of the Schroedinger operator $E=L+\alpha V$, with respect to the smallest eigenvalue. Thus, as in the case of Laplacian Eigenmaps, we skip the smallest eigenvectors and take the $n$ subsequent ones. 
\end{remark}

\begin{figure}
\centering
\subfigure[(left) Original arc in three-dimensional space. (right) Laplacian Eigenmaps perfectly recovers the two-dimensional structure.]{
\includegraphics[width=0.15\textwidth]{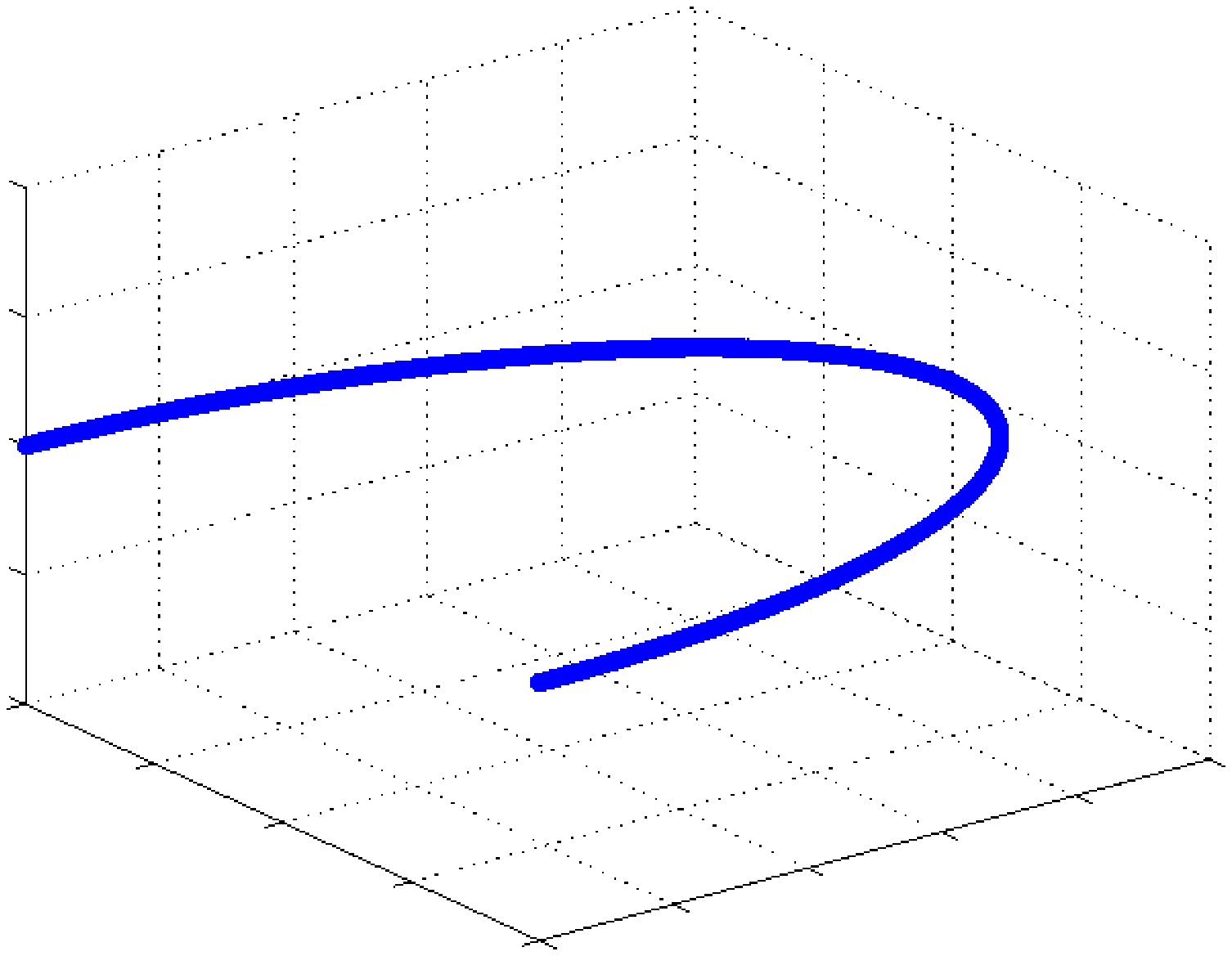}
\includegraphics[width=0.115\textwidth]{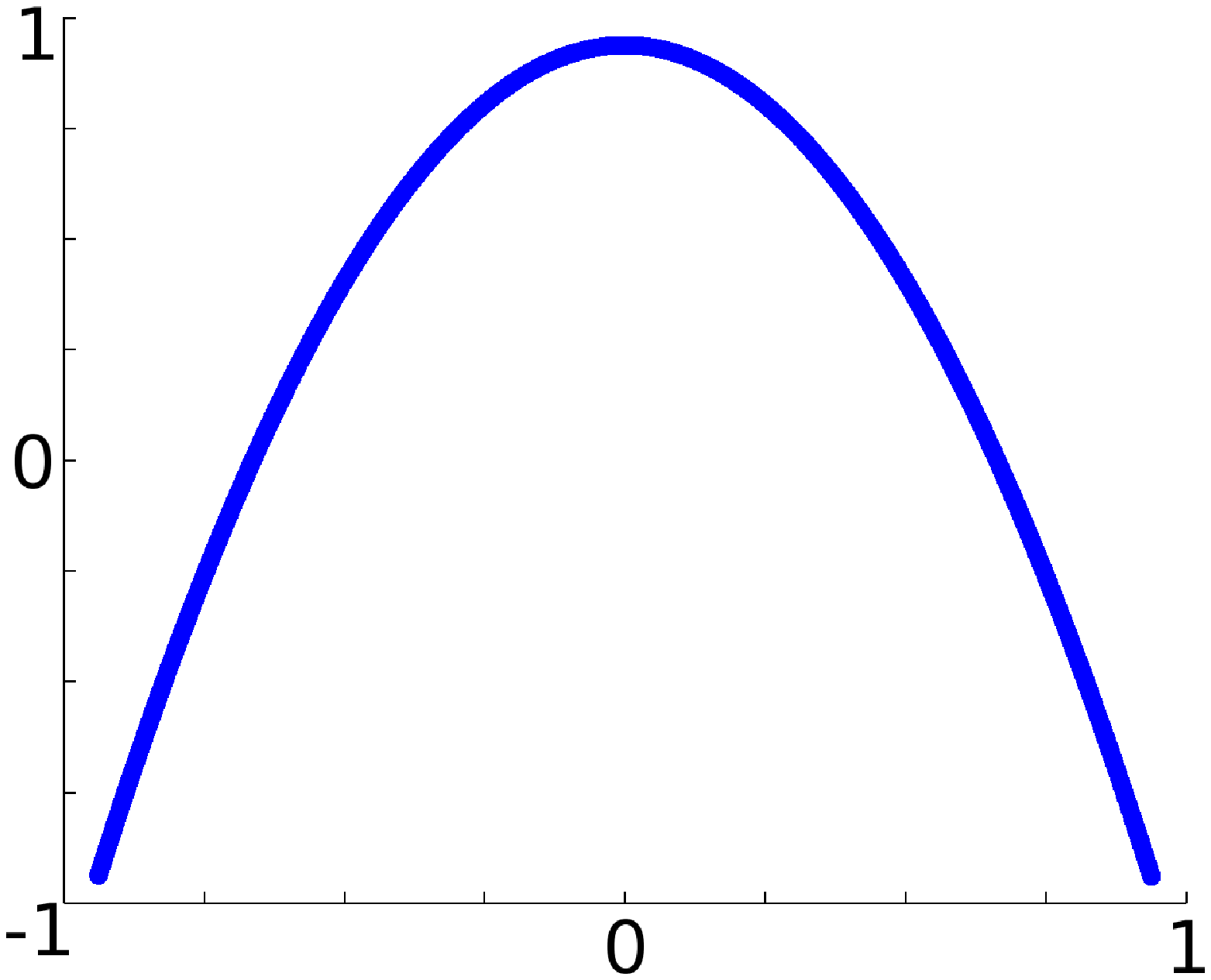}
}

\subfigure[The Schroedinger Eigenmaps with diagonal potential $V=\diag(0,\ldots,0,1,0,\ldots,0)$ only acting in one point $y_{i_0}$ in the middle of the arc for $\alpha=0.05, 0.1, 0.5, 5$. This point is pushed to zero. Since $V$ is 'discontinuous', the dimension reduced arc gets more and more discontinuous until $y_{i_0}$ will become zero for sufficiently large $\alpha$. Then the arc does not deform any more.]{
\includegraphics[width=0.115\textwidth]{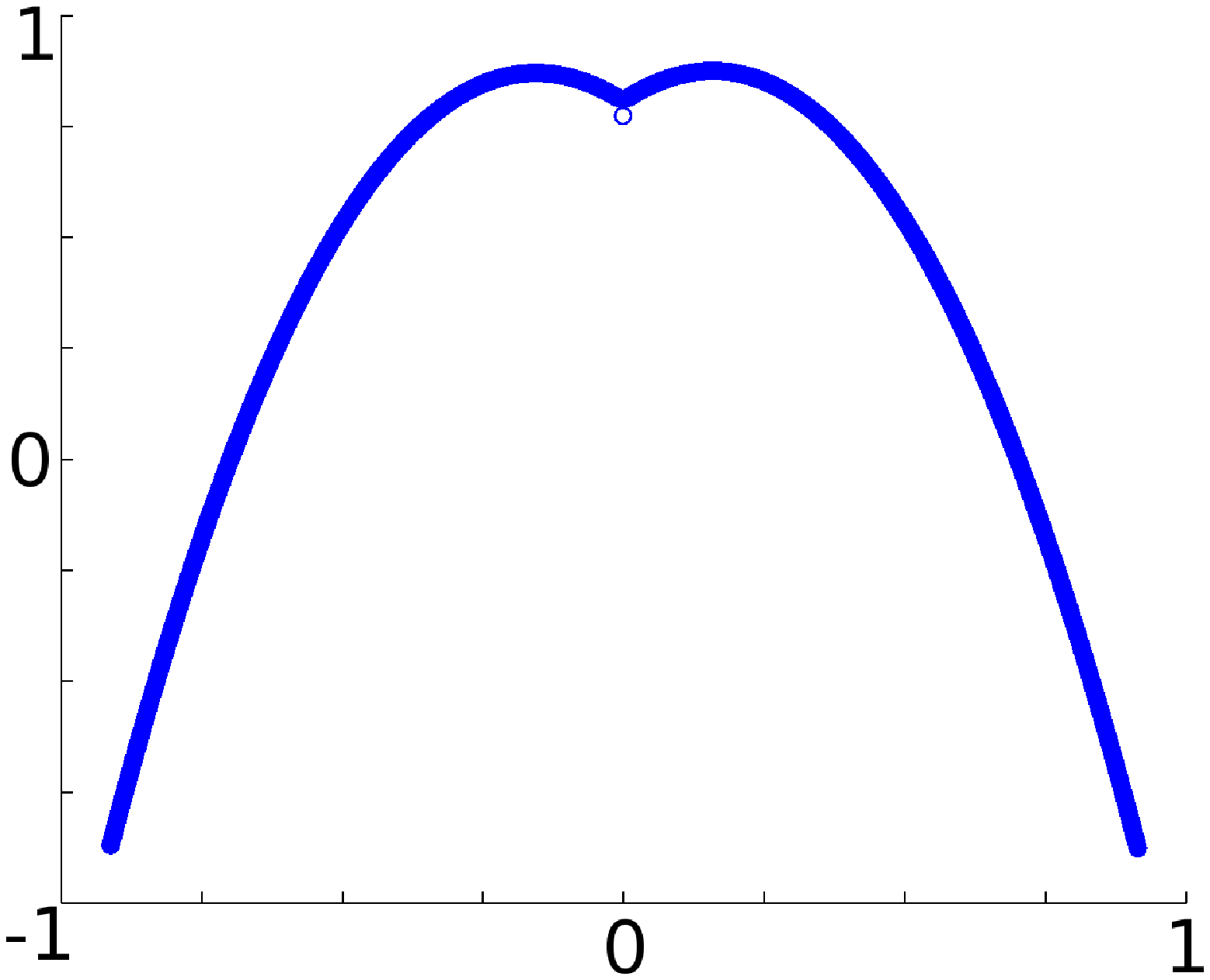}
\!\includegraphics[width=0.115\textwidth]{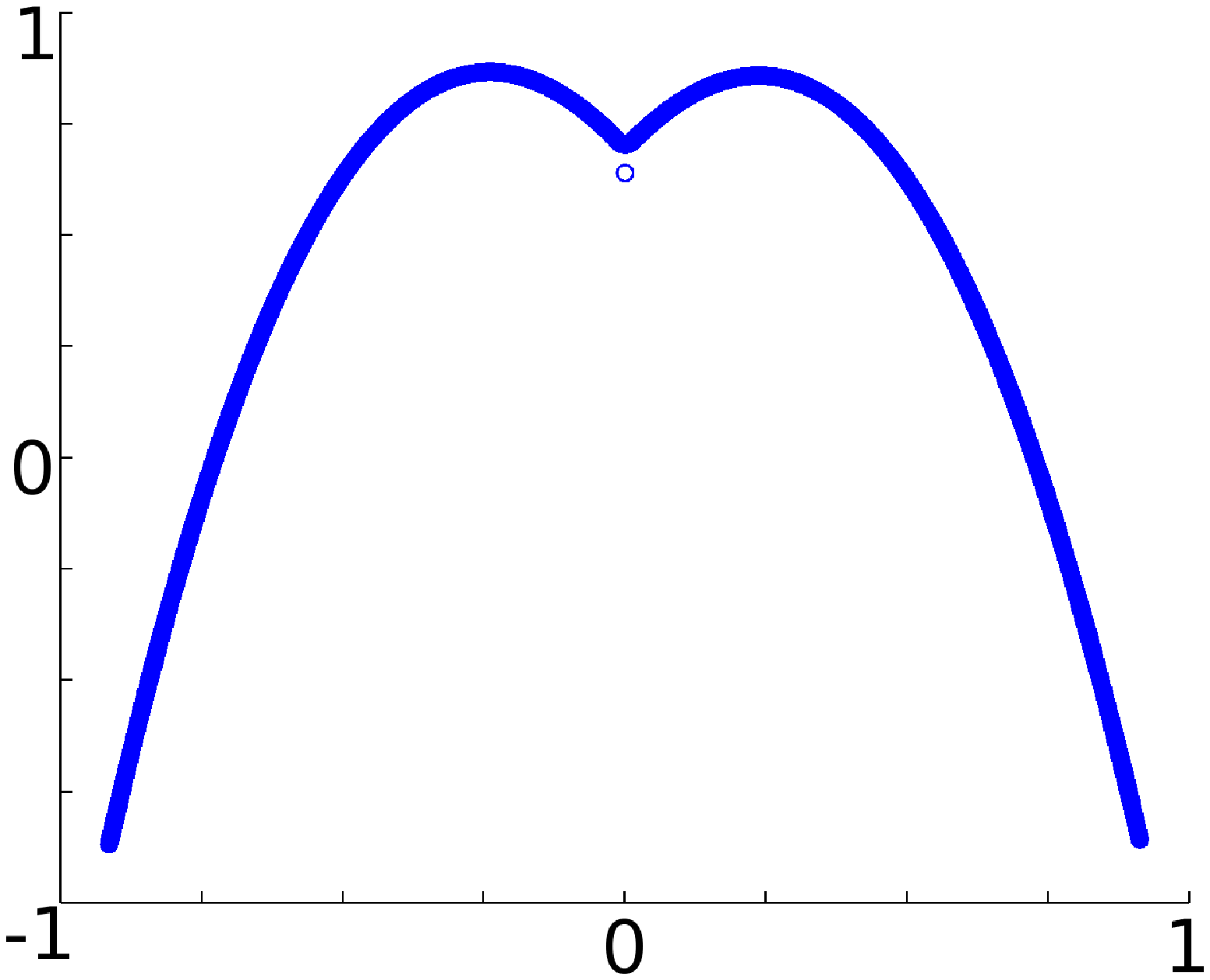}
\!\includegraphics[width=0.115\textwidth]{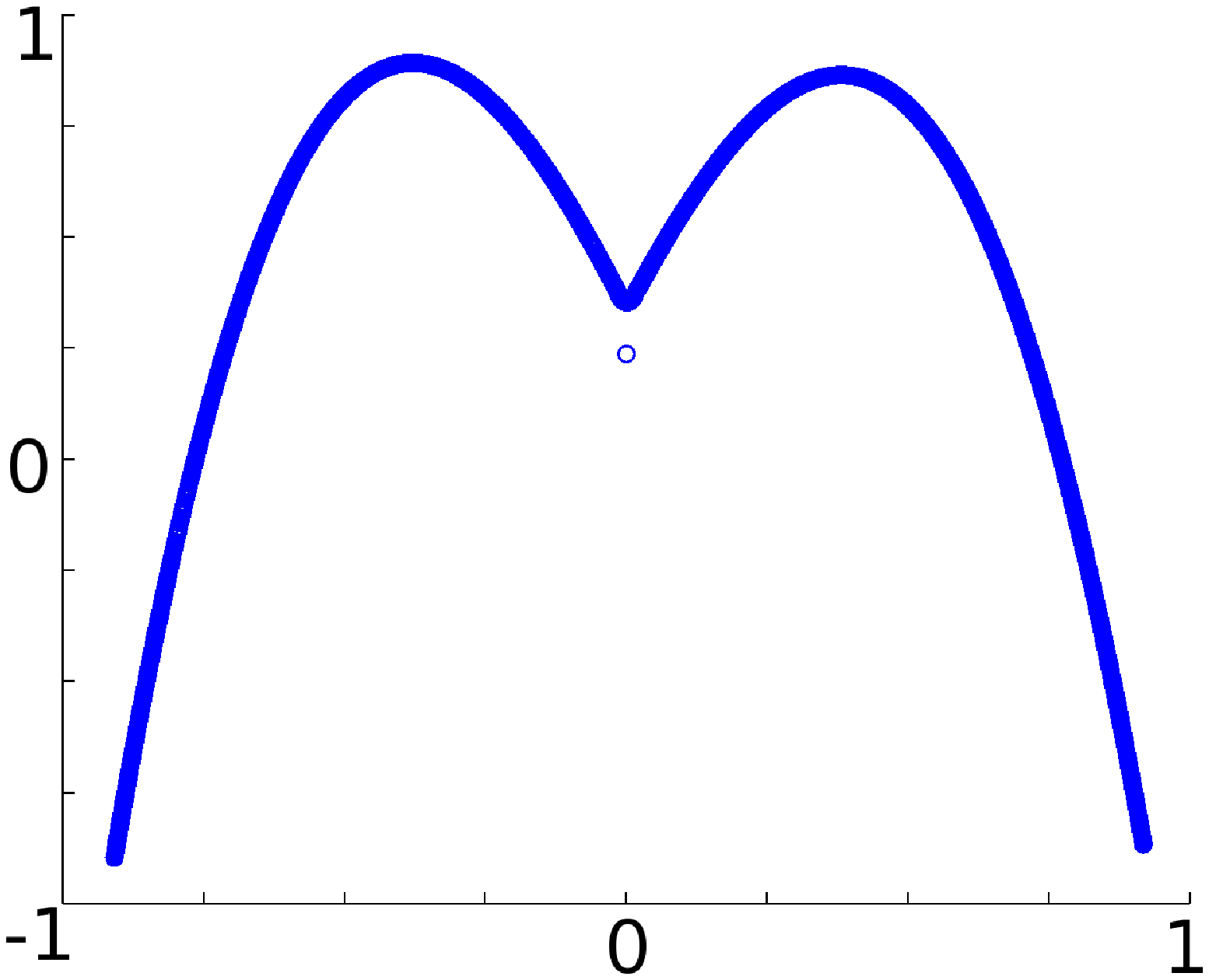}
\!\includegraphics[width=0.115\textwidth]{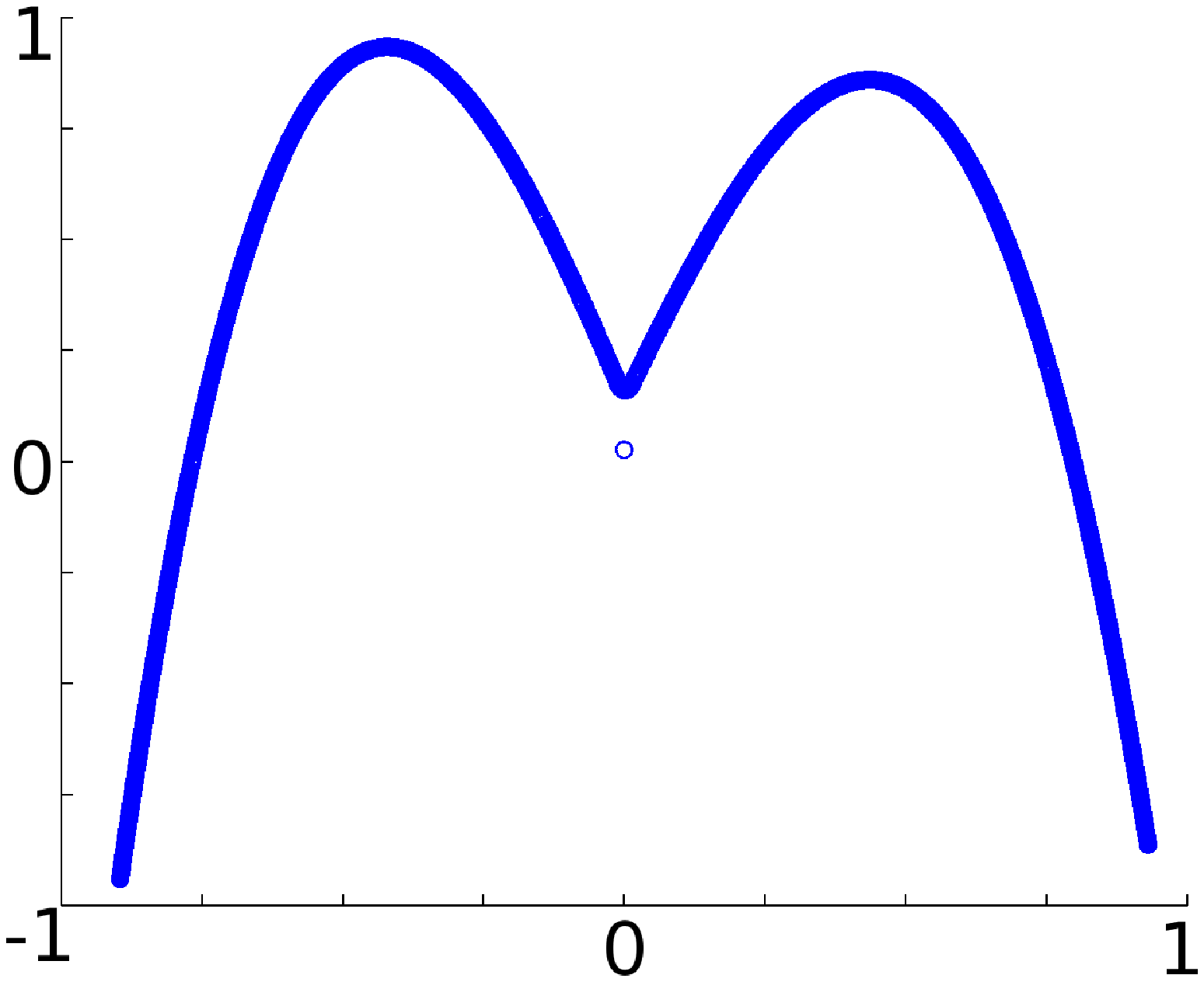}
}

\subfigure[By labeling the end points of the arc within the nondiagonal potential $V^{(i,j)}$, for $\alpha=0.01, 0.05, 0.1,1$, we are able to control the dimension reduction such that we obtain an almost perfect circle (up to a small discontinuity).]{
\includegraphics[width=0.115\textwidth]{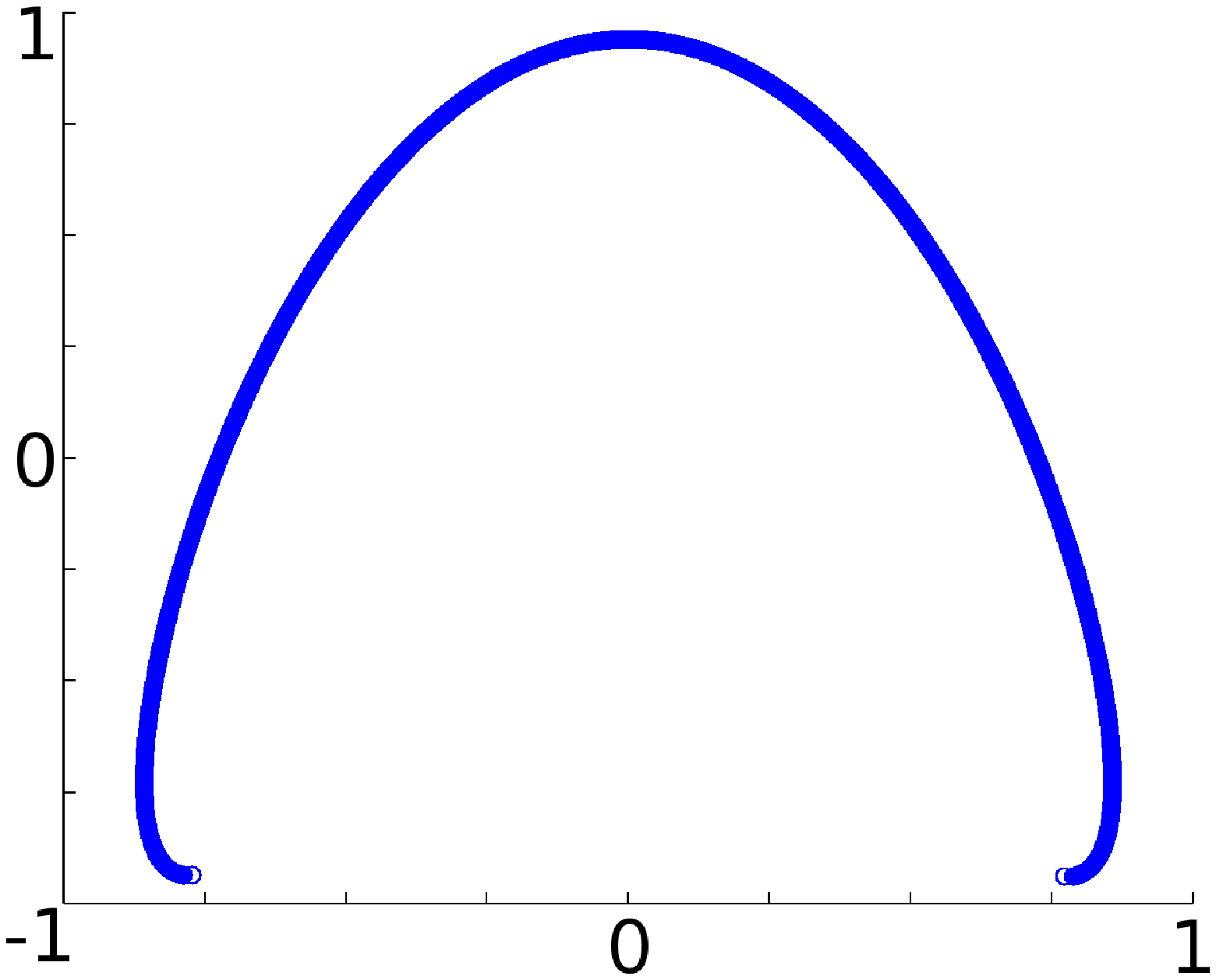}
\!\includegraphics[width=0.115\textwidth]{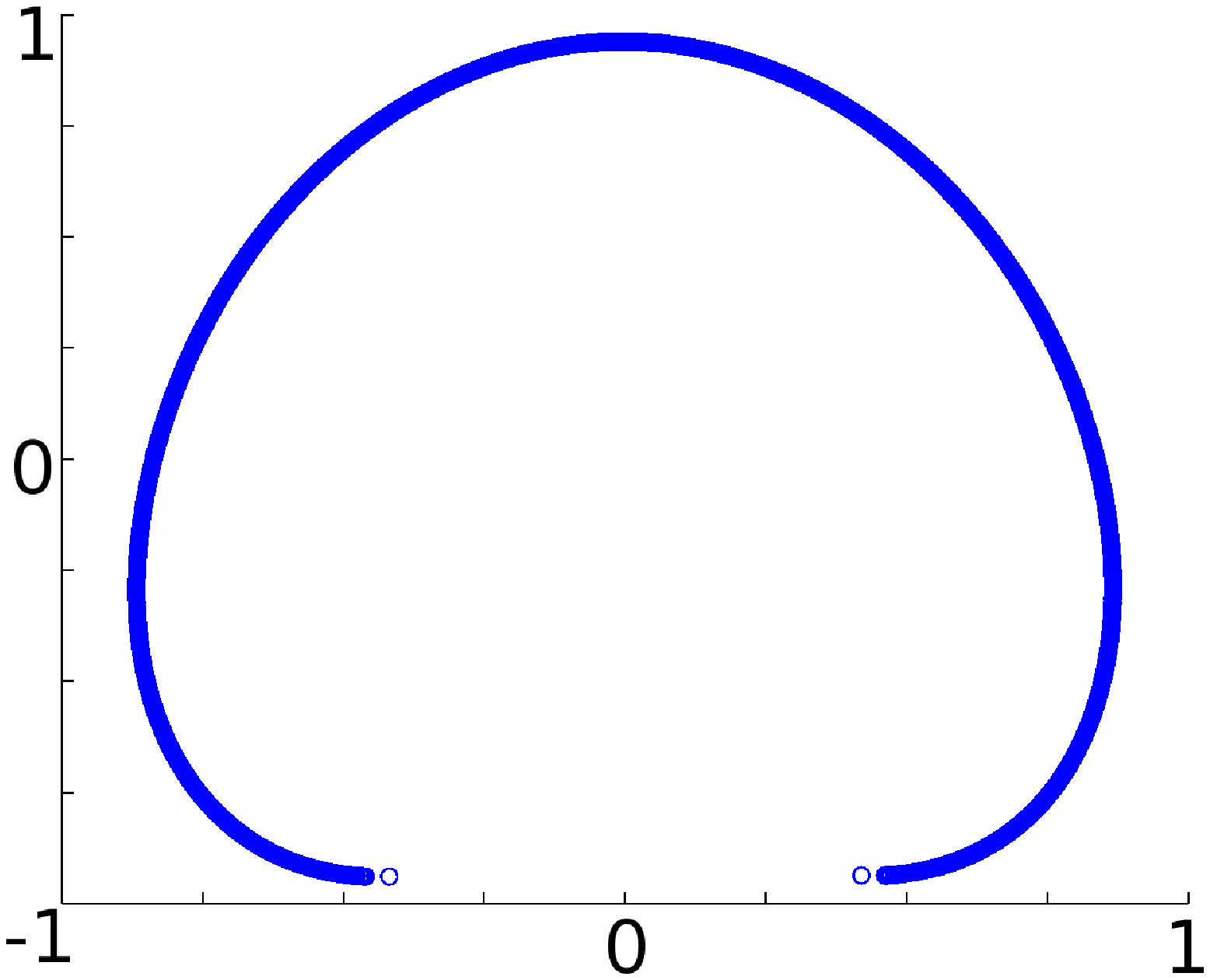}
\!\includegraphics[width=0.115\textwidth]{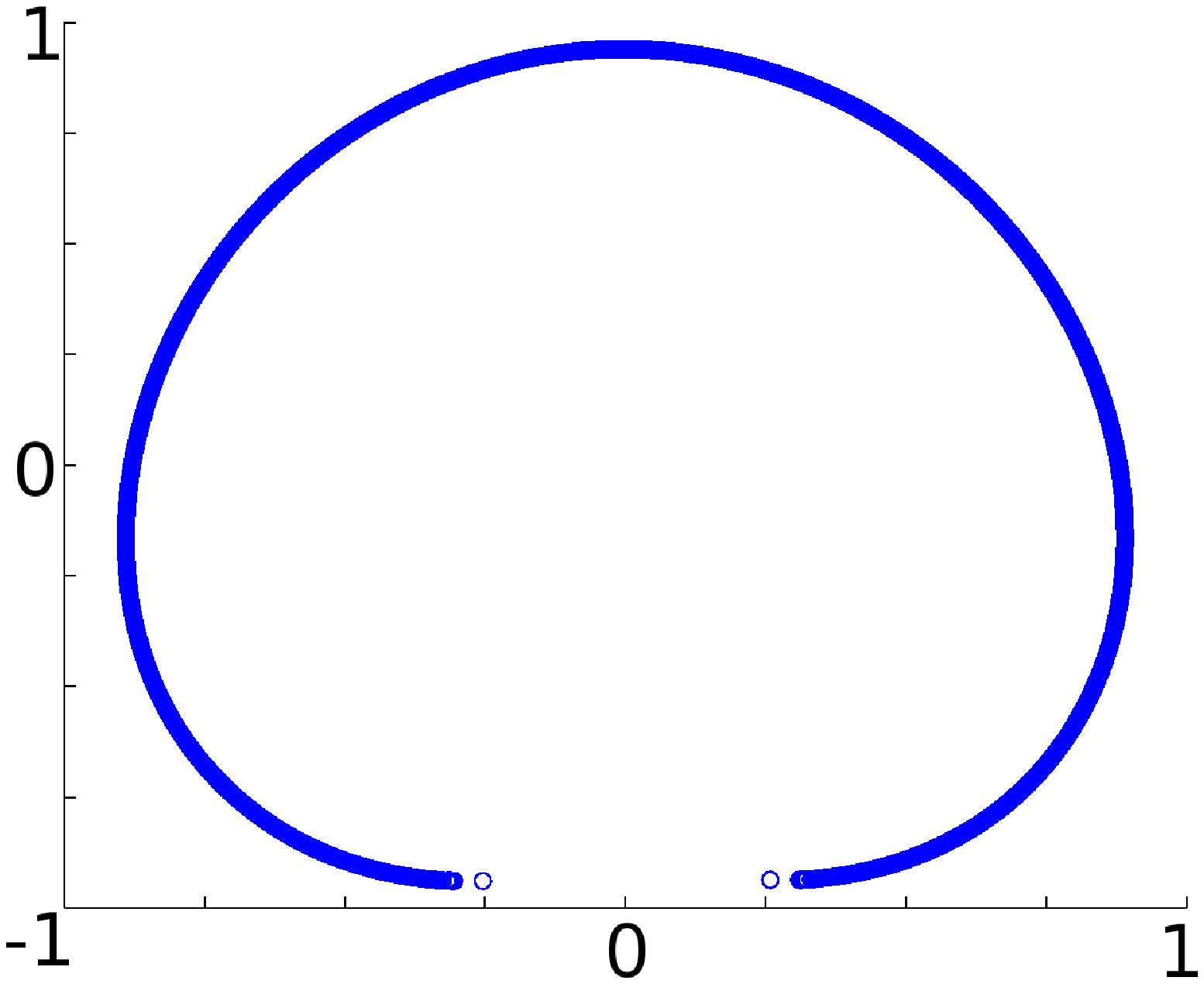}
\!\includegraphics[width=0.115\textwidth]{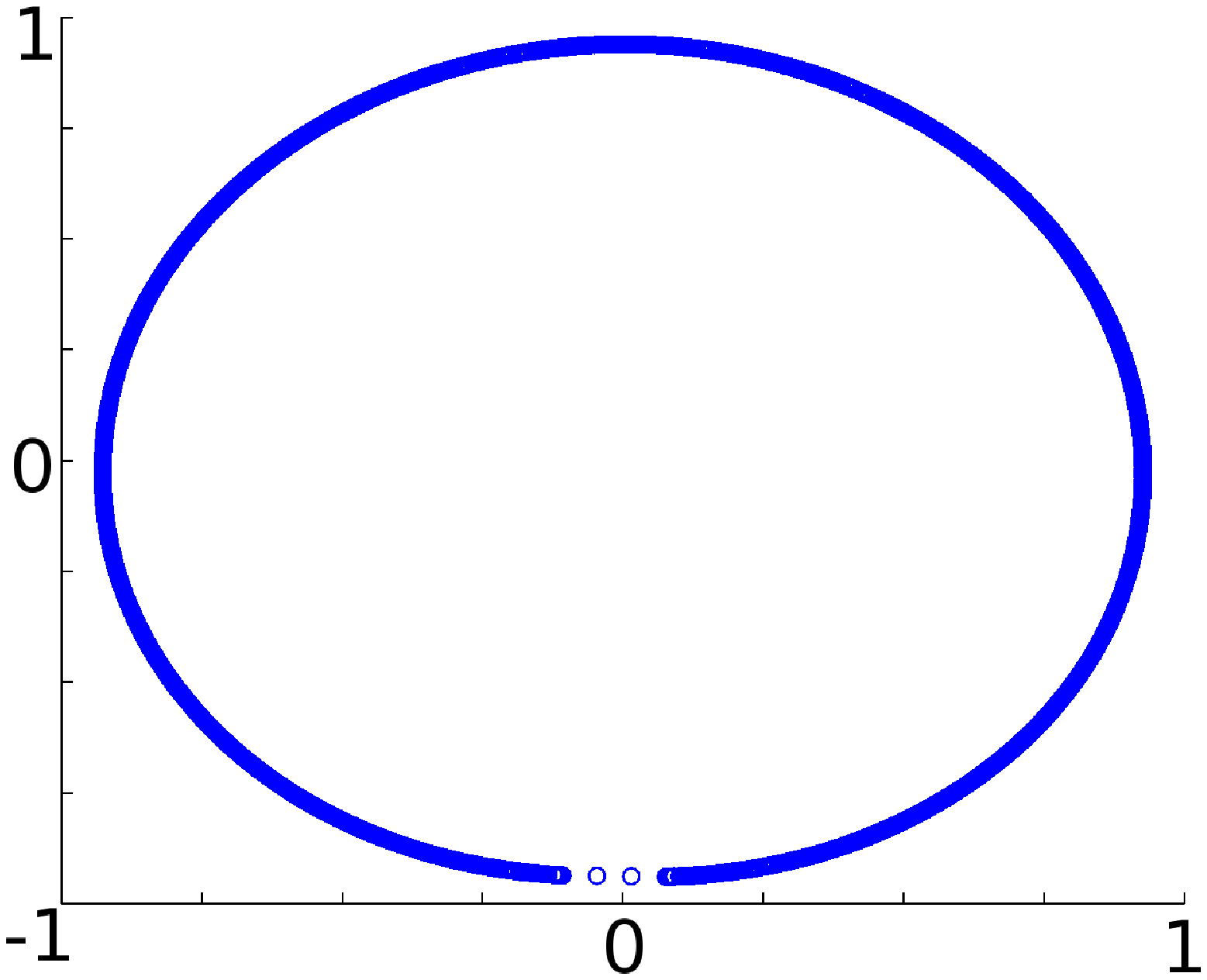}
}
\caption{Schroedinger  Eigenmaps applied to an arc}\label{figure:arc}
\end{figure}

To demonstrate the use of the barrier potential, SE is applied to a three-dimensional arc in Figure \ref{figure:arc}.    

\begin{remark}
For the purpose of sequentially identifying points, we propose to use the potential 
\begin{equation}\label{eq:DCT}
V=\sum_{i=1}^{m-1}V^{(i,i+1)},
\end{equation}
which penalizes $L$ by $\sum_{i=1}^{m-1}\|y_i-y_{i+1}\|^2$. Note that the eigenvectors of this matrix are the basis vectors of the discrete cosine transform of type II.
\end{remark}



\subsection{Schroedinger Eigenmaps For Classification}\label{section classification}
In this section we discuss the standard Vector Angle Classification (VAC) in combination with Schroedinger Eigenmaps. In this regard, suppose we have $s$ pairwise different seed vectors $a_1,\ldots,a_s\in\R^n$, which represent the centers of each class. The class $C_i$ consists of $y^{(\alpha)}_i$s whose angle is closest to the specific center vector as long as the angle is smaller than some tightness parameter $\delta_i$. We additionally apply a threshold parameter $\delta$ to refuse $y^{(\alpha)}_i$s that have small norm, which by themself form another class. 

According to Corollary \ref{corollary:direct}, SE can be applied to ``force" processed pixels to zero at the rate $1/\sqrt{\alpha}$ and, hence, below any threshold $\delta>0$ for sufficiently large $\alpha$. A practitioner can use labels in the form of a diagonal potential to separate a particular region of an image, see Section \ref{RI}. On the other hand, different spatial regions can be identified with each other by labelling them with non-diagonal potentials $V^{(i,j)}$, cf., \eqref{eq:DCT}. This is useful to classify pathological areas as one single class despite the fact that intrinsic intra-class variations can be large.

As it was mentioned in the introduction, a very considerable effort has been aimed at developing partially labeled classification schemes based on the Laplacian Eigenmaps algorithm,  on related graph random walks, or on Diffusion Maps. The classification scheme based on Schroedinger Eigenmaps with partially labeled data is novel. When compared to ``warping" and function adapted diffusion processes \cite{Coifman:2005aa} \cite{Szlam:2008kx}, Schroedinger Eigenmaps result in a more general class of matrices (operators), by allowing modifications to the diagonal as well as to the off-diagonal terms. These diagonal terms, in turn, are associated with locations of the barrier potential which affects  and steers the diffusion process. On the other hand, Schroedinger Eigenmaps classification is a joint optimization process for labeled and unlabeled data, as opposed to classical semi-supervised algorithms  \cite{Belkin:2002ys}, \cite{Belkin:2004vn}.

%
%



\section{Biomedical Applications}\label{section experiments}
The performance of Schroedinger Eigenmaps on three standard medical datasets is compared with linear and Gaussian kernel Support Vector Machines (SVM). After having verified the usefulness of Schroedinger Eigenmaps, we apply them to analyze new problems in genetic expression analysis and retinal multispectral imaging. 

%



\subsection{Three standard bio-medical datasets}\label{sec:two standard}
All $3$ standard datasets are obtained from the UCI Machine Learning Repository. Training points are randomly selected. The potential is designed on the training data and SE are computed on the entire dataset so that VAC with threshold leads to the final class separation. 

After removing missing values, the \emph{Wisconsin Breast Cancer Database} (WBCD) \cite{Wolberg:1990uq} contains $683$ patterns with $9$ attributes and is grouped into two classes, benign and malignant. The diagonal potential is used on the benign class in the training data. The processed vectors that fall below a threshold  form the predicted benign class on the entire dataset. 

The \emph{Cleveland Heart Disease Database} (CHDD) \cite{Detrano:1989fk} contains $297$ patterns with $13$ attributes and is grouped into presence of heart disease (values 1,2,3,4) and absence (value 0). We removed $6$ patterns due to missing values. Following the focus in the literature, we aim to separate absence from heart disease. We label absence in the training data using the diagonal potential and identify heart disease classes $1$-$4$ by applying the nondiagonal potential. The predicted absence class is formed by the vectors that fall below the threshold.  

The \emph{Mammographic Mass Data Set} (MMD) \cite{Elter:2007fk} contains $961$ instances and $5$ attributes that are grouped into benign and malignant. Note that we removed $131$ patterns due to missing values and suppressed the first attribute as it was a numerical assessment of the range between benign and malignant of a  double-review process by physicians.

In our computations of SE, we keep $6$ eigenvectors, and the number of nearest neighbors varies between $6$ and $20$, the weight parameter $\alpha$ in the potential, and the threshold parameter are optimized for minimizing the error rates. We compare our method with  Laplacian Eigenmaps (same parameters as for SE) and with SVM. The Matlab implementation of SVM is used with a linear and a Gaussian kernel function, and the separating hyperplane is found from least squares or a one-norm with soft-margin svm; the best result is taken. The parameter in the SVM Gaussian kernel corresponds to $\sigma$ in Schroedinger Eigenmaps. After optimization of $\sigma$ in SVM for each dataset separately, it appeared consistent that best results for both, SVM and SE, were obtained by choosing $\sigma=1/2,1,2$ in SE and $2\sigma$ in SVM for the datasets WBCD, CHDD, and MMD, respectively.

Error rates for large training data yield information on the nonlinear data structure. For instance, as the error of the linear SVM applied to WBCD with $600$ training points does not vanish, the data cannot be separated in a linear fashion, see Table \ref{table:standard}. 

The comparison to Laplacian Eigenmaps (same parameters as for Schroedinger Eigenmaps) shows how the potential can improve the classification process.  If there are only few training data, then Schroedinger Eigenmaps lead to the smallest error rates among all methods, most considerably for CHDD and MMD. When increasing the number of training points, the Gaussian kernel SVM yields the best results except for MMD. Typical bio-medical data allow for only few reliable training data and in this situation Schroedinger Eigenmaps appear most useful. 

\begin{table}
\begin{tabular}{|c|c|c|c|c|}
\hline
 dataset, \#training data &LE & SVM lin. & SVM Gauss.  &  SE\\ 
\hline
WBCD, $40$ & $14 \%$ & $5\%$  & $5\%$ & $4\%$\\
WBCD, $100$ & $9\%$ & $4\%$  & $4\%$ & $3\%$\\
WBCD, $200$ & -- & $4\%$  & $3\%$ & $3\%$\\
WBCD, $400$ & -- & $4\%$  & $2\%$ & $3\%$\\
WBCD, $600$ & -- & $4\%$  & $2\%$ & $3\%$\\
\hline
CHDD, $40$ & $42\%$ & $21\%$ & $19\%$ & $15\%$\\
CHDD, $100$ & -- & $17\%$ & $15\%$ & $12\%$\\
CHDD, $200$ & -- & $16\%$ & $11\%$ & $12\%$\\
CHDD, $297$ & -- & $15\%$ & $9\%$ & $11\%$\\
\hline
MMD, $20$ & $45 \%$ & $24\%$  & $24\%$ & $22\%$\\
MMD, $30$ & $40 \%$ & $22\%$  & $22\%$ & $21\%$\\
MMD, $40$ & $38 \%$ & $22\%$  & $21\%$ & $20\%$\\
MMD, $100$ & -- & $21\%$  & $20\%$ & $20\%$\\
MMD, $200$ & --& $20\%$  & $19\%$ & $20\%$\\
MMD, $800$ & -- & $20\%$  & $18\%$ & $20\%$\\

\hline

\end{tabular}
\caption{Error rates as incorrectly classified samples / total data are averaged over $100$ instances for each method. LE is combined with VAC, where training data are used as seeds that always end up in the correct class. Therefore, the LE outcome for large training data is not due to the LE method
}\label{table:standard}
\end{table}


\subsection{Applications to new biomedical data}
\label{RI}

Up to this point we have compared Schroedinger Eigenmaps with other methods for semi-supervised classification. We believe the strength of our proposed technique is threefold: good performance with few labeled points, a global
approach yielding good local discrimination, and ability to correctly classify in the presence of non-linear mixing. In this section we shall illustrate some of these points with two examples: a
classification problem in medical imaging, and a clustering problem in genetic
analysis. At the same time, these two examples give additional evidence for the results of the first part of our paper.

\subsubsection{Analysis of clinical, retinal images}
First, we analyze a problem in retinal imaging, where our goal is to detect and differentiate between two classes of chemical mixtures in the human retina. This problem is of interest to the medical community, as it appears in studies of age-related macular degeneration (AMD) -- the leading cause of blindness among the elderly population in the developed world, see  \cite{Chew:2009aa,Chew:2009ab,Coleman:2008aa,Meyers:2004ab}. 
The balance between these two chemical mixtures, which we want to detect, can be used to determine the stage of the disease and, in turn, the proper course of action \cite{Hageman:2001vn,Haimovici:2001uq,Wang:2010kx}. 
%
There is demand in the medical community for automated analysis tools that enable detection and classification while, at the same time, allowing for expert input \cite{Ehler:2011aa,Kainerstorfer:2010ab,Sbeh:2001fk}.

We shall apply our Schroedinger Eigenmaps technique, with the locations of the potential chosen by medical experts.
This example further illustrates that relatively few (less than 0.1\%)
labels may lead to good classification results. An added difficulty in this scenario is that, due to limitations of the imaging  device, the spectral information about the chemicals of interest is always 
nonlinearly mixed with various other responses, e.g., from layers just beneath 
the retina, or from different illumination patterns. As such, most of the mixing 
models that rely on presence of pure pixels in the image fail to work.

For each patient, 4 excitation with 2 emission filters and trifold imaging 
lead to $24$ images ($400\times 400$ pixels) that are aligned by applying the commercial software i2k Align. One pixel, therefore, has $24$ entries in $z$-direction, so that the acquired dataset are $160000$ vectors in $24$-dimensional space. The earliest clinical signs of AMD are bright spots in retinal reflection images, see Fig.~\ref{labelle}(a), referred to as \emph{drusen} \cite{Bird:1995aa}.  We have applied Schroedinger Eigenmaps with diagonal potential acting on a small predefined drusen area ($50$ pixels) that was identified by medical experts. The number of nearest neighbors was chosen to be $14$, $\sigma=1$, $\alpha$ in the potential is $10$ times the average norm of the data vectors, and we used $15$ eigenvectors. We obtain a drusen map of all pixel vectors whose Schroedinger Eigenmaps coordinates fall below a threshold that was adjusted through visual inspection, cf.~Fig.~\ref{labelle}(b).



It was suggested in \cite{Delori:2000fk}, that drusen centers may separate from outer drusen areas. To distinguish the two drusen subclasses, we additionally form a second potential on a drusen center ($75$ pixels) that identifies pixels and run SE simultaneously with both potentials activated. We obtain two subclasses of drusen, one by pixels whose Schroedinger Eigenmaps coordinates fall below the threshold, the other one is derived from taking the average over identified pixel vectors as a seed in VAC. 
Threshold and class tightness parameter are optimized through visual inspection,  
see Fig.~\ref{labelle}(c). The same label locations are used for linear and Gaussian SVM that are applied to the detected drusen in Fig.~\ref{labelle}(b) for further subclassification, see Fig.~\ref{labelle}(e,f).

\newlength{\imLength}
\setlength{\imLength}{.2\textwidth}

\begin{figure}
\centering
\subfigure[drusen appear as bright spots; one is marked by the green square; detection and further subclassification is needed]{
\includegraphics[width=\imLength,height=\imLength]{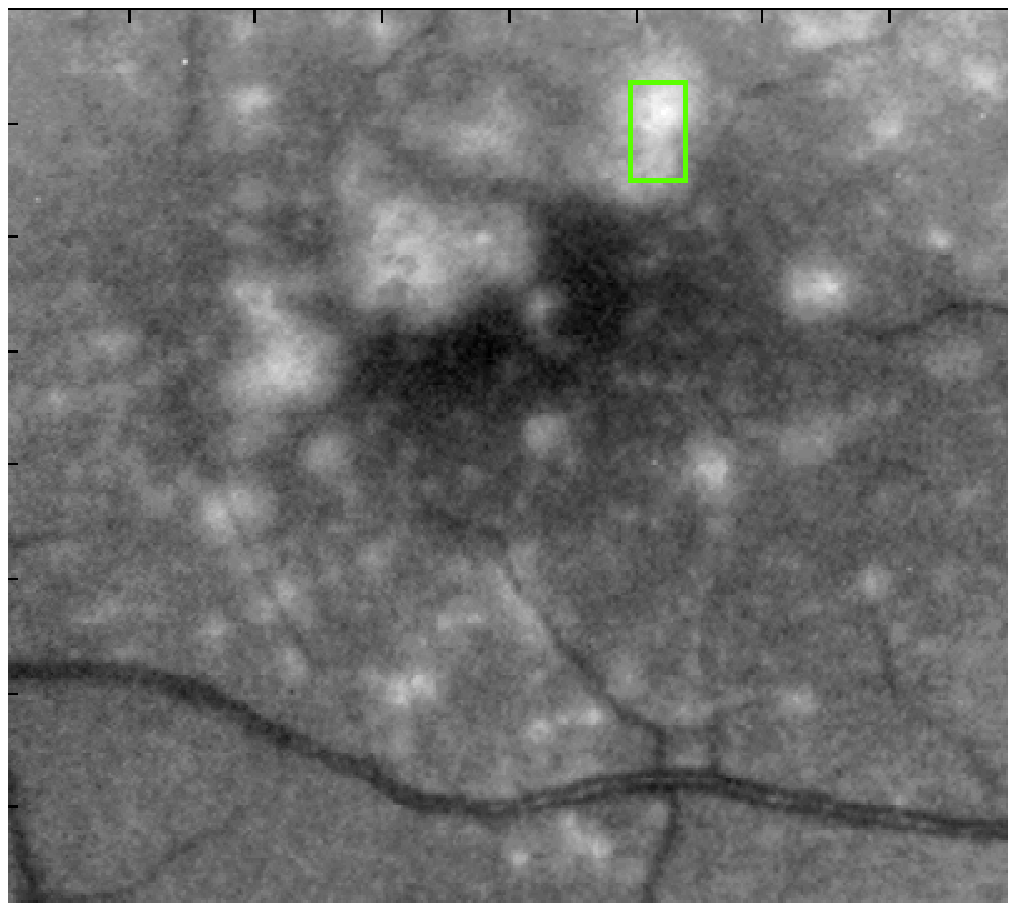}}
\hspace{2ex}
\subfigure[only a small drusen region ($50$ pixels) in the upper left corner was penalized in the barrier potential providing one class of drusen]{
\frame{\includegraphics[width=\imLength,height=\imLength]{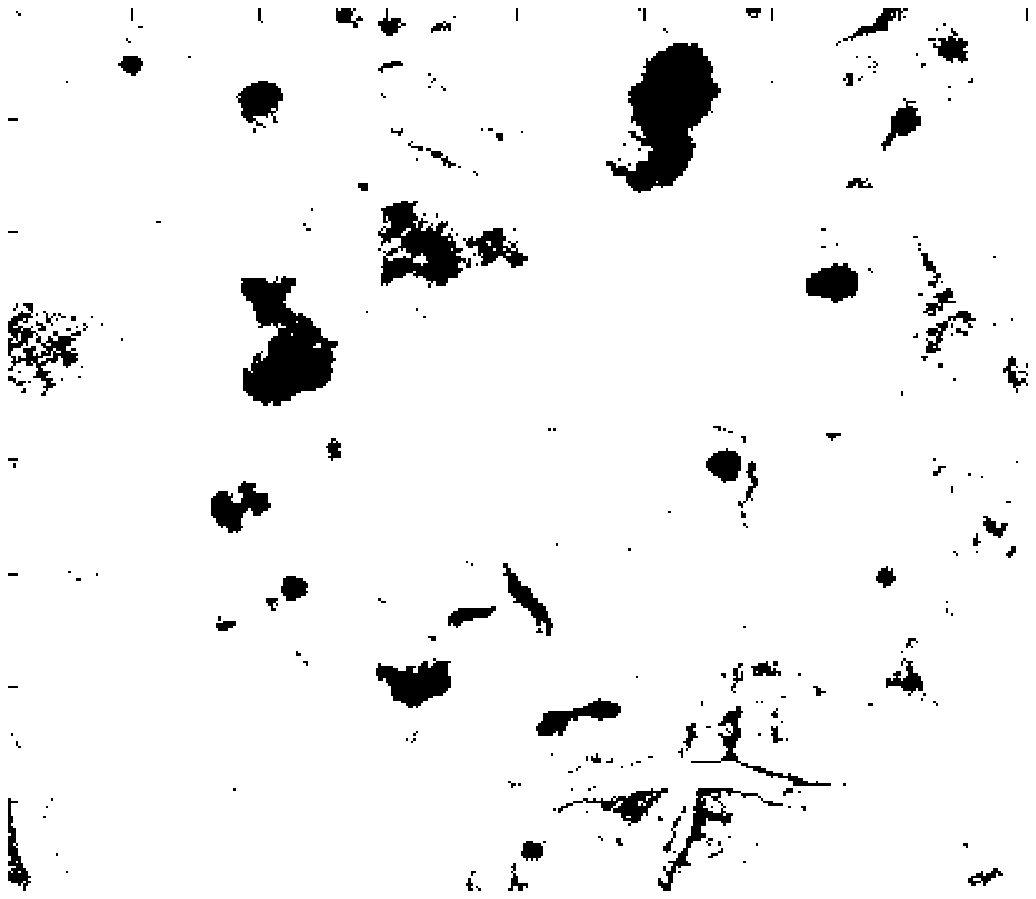}}}

\hspace{1ex}
\subfigure[blue area ($75$ pixels) was additionally  identified by SE (non-diag.~matrix); VAC with threshold yields $2$ well-se\-pa\-rated drusen classes (black/red). Consistent with \cite{Delori:2000fk}, we separate drusen centers from outer drusen segments]{
\frame{\includegraphics[width=\imLength,height=\imLength]{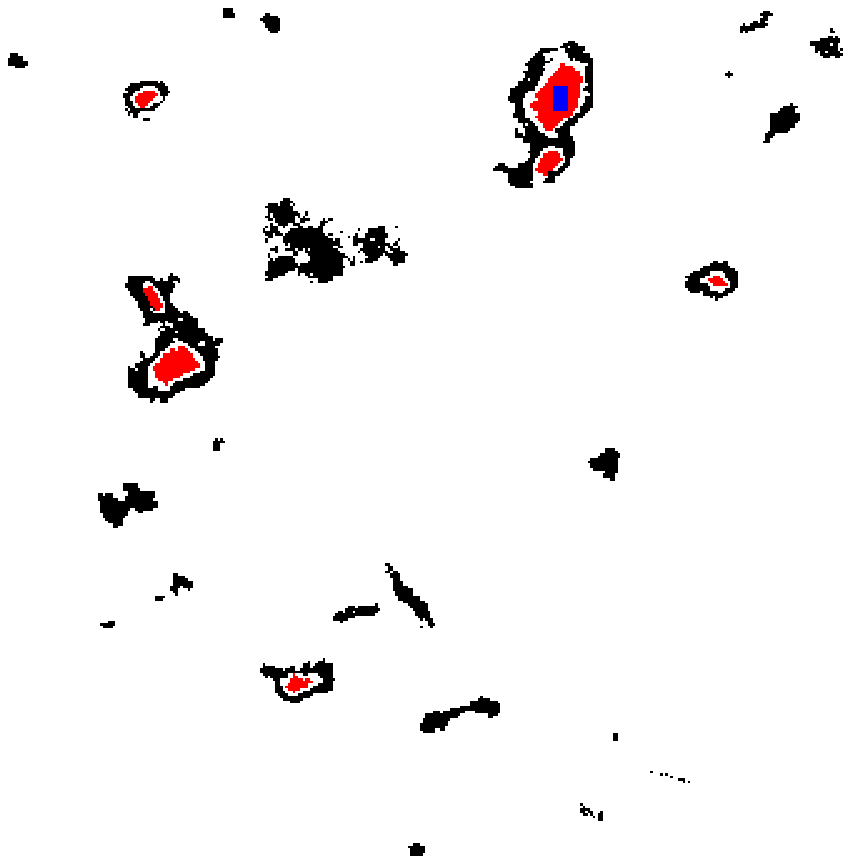}}}
\hspace{2ex}
\subfigure[zoomed onto the marked drusen in (a); red lines show LRW segmentation; drusen center and outer drusen area were separated by using $3$ sets of labels; drusen center (blue), drusen outer area (yellow), background (green)]{
\hspace{4.5ex}
\includegraphics[height=\imLength]{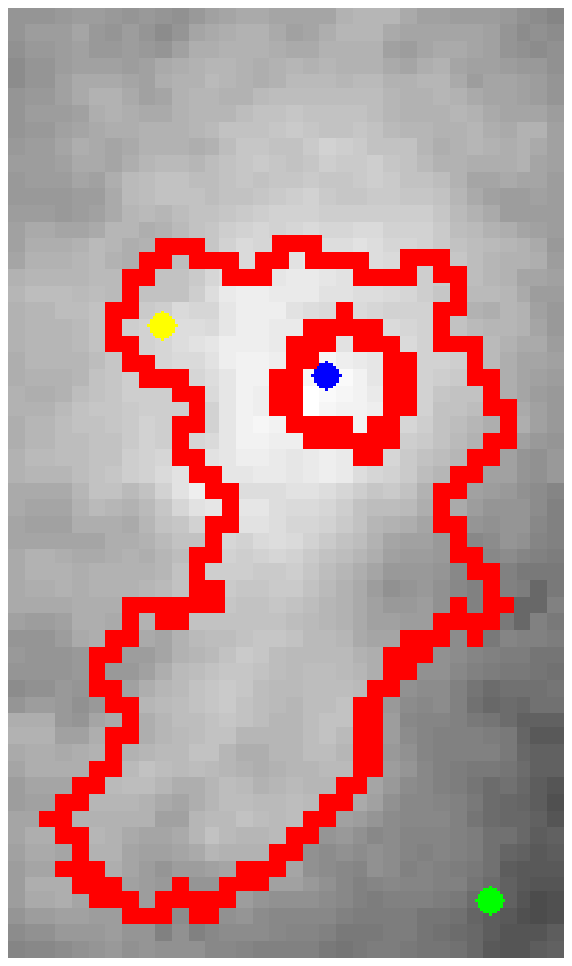}
\hspace{4.5ex}
}

\subfigure[The class identified in (b) is further subclassified by running linear SVM on this subset with labels on the drusen center]{
\frame{\includegraphics[width=\imLength,height=\imLength]{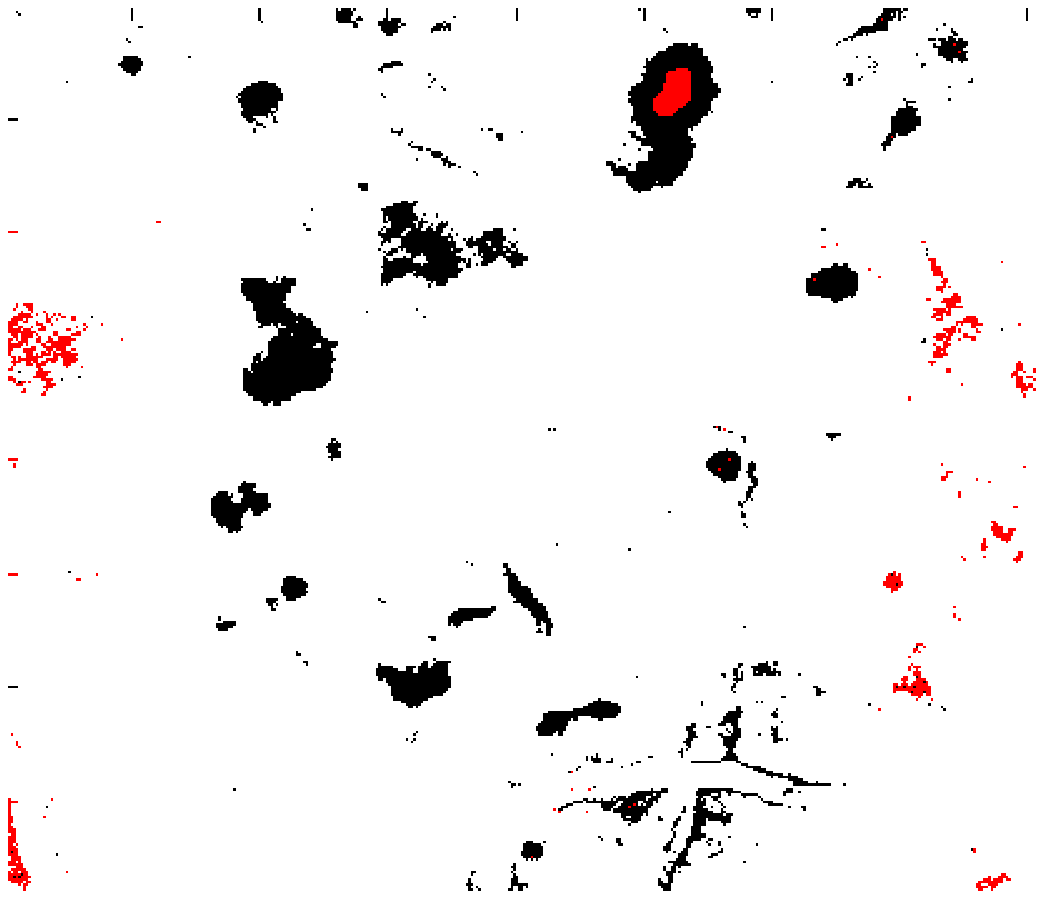}}}
\hspace{2ex}
\subfigure[Gaussian SVM with labeled data as in (e) applied to the class in (b)]{
\frame{\includegraphics[width=\imLength,height=\imLength]{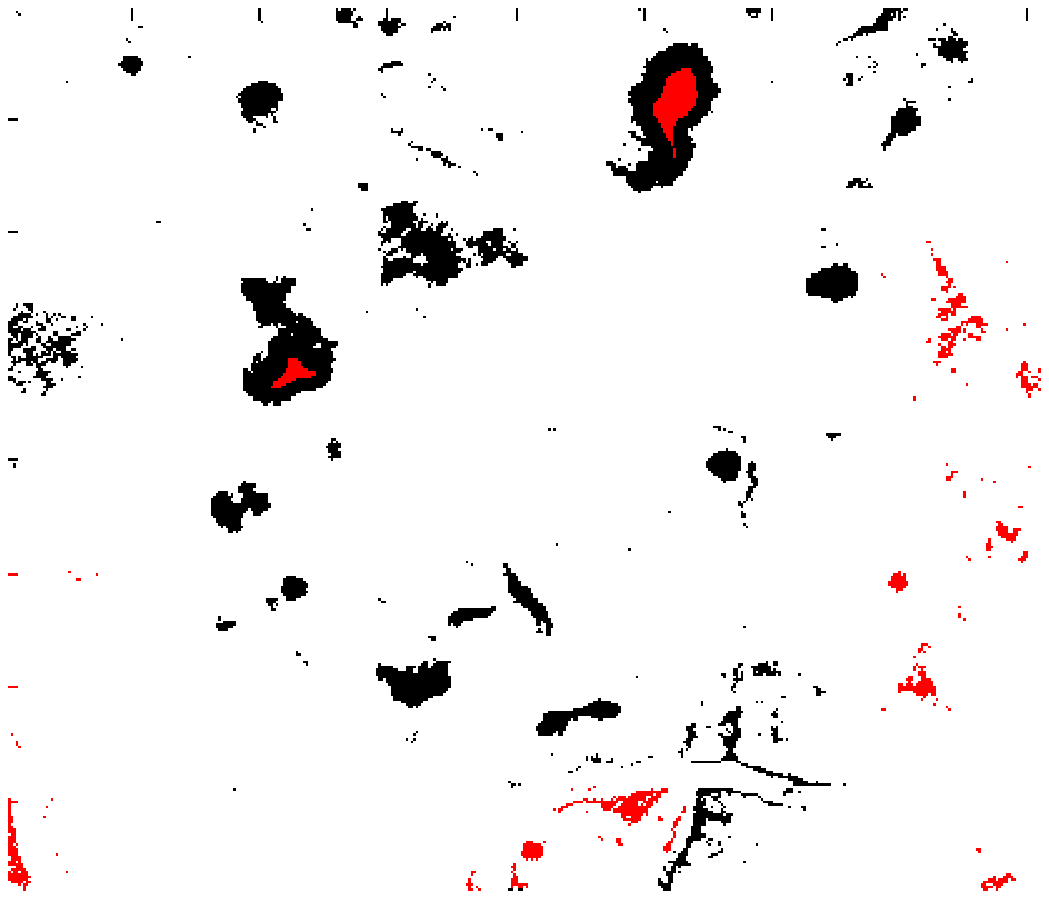}}}

\caption{Classification with Schroedinger Eigenmaps, LRW verification, and  comparisons to SVM based on multi-spectral retinal image sets}\label{labelle}


\end{figure}


In a single drusen, members of the same class are spatially connected, so that we can apply the segmentation tool Laplacian Random Walk (LRW) in \cite{Grady:2006aa}, confirming the drusen subclassification, cf.~Fig.~\ref{labelle}(d).  SE appears well-suited to incorporate spatially unconnected regions, going beyond image segmentation.

\subsubsection{Analysis of gene expressions}
Recent efforts in biology focus on the understanding of genetic datasets based on complex gene interactions, rather than analyzing genes individually. This shift of perspectives increases the importance of small local variations, which may be irrelevant from a purely global perspective. 
Therefore, analysis and classification of time series gene expressions require the balance between local and global variations. 

Hierarchical clustering  in \cite{Brown:2009ab} and Laplacian Eigenmaps  in \cite{Ehler:2010ab} was used to tentatively analyze a collection of $8316$ microarray gene expressions at $8$ time-points related to eye morphology in mice, see \cite{Bonner:1997aa,Zeeberg:2003aa} for further background. The goal of those studies was to identify new gene clusters within this dataset that are associated to eye development. 

We used a barrier potential in \cite{Ehler:2010af} to label certain DNA specific proteins (transcription factors) and found a set of new genes related to eye morphogenesis, which were not discovered by traditional hierachichal techniques. Thus, even a preliminary application of the barrier potential for data labeling proved useful in \cite{Ehler:2010af} to add complementary expert data enabling improvements in gene clustering.

\section{Conclusions}
We have introduced a generalization of Laplacian Eigenmaps to allow for expert data in form of a potential on a data-dependent graph. The properties of our novel Schroedinger Eigenmaps are studied, and we illustrate their action on both, artificial and standard medical data. At this stage these are only illustrations, and providing mathematical formalism behind these results is part of our on-going and future work on this subject. It appears that our method is robust and particularly useful if the data quality is low and there are only few training data, a typical situation in bio-medical applications. In addition, our scheme is successfully applied to the analysis of retinal multispectral images. We classify drusen based on few training pixels that were labeled by a physician. Our results suggest that Schroedinger Eigenmaps are useful in high-dimensional bio-medical data analysis when limited expert knowledge is available.

\appendices

\ifCLASSOPTIONcompsoc
  
  
  \section*{Acknowledgments}
\else
 
 
  \section*{Acknowledgment}
\fi

This work was supported by the Intramural Research Program of NICHD/NIH, by NSF (CBET0854233), by NGA (HM15820810009), by ONR (N000140910144), and by NIH/DFG (EH 405/1-1/575910). The authors also express their gratitude to Drs. R.~F.~Bonner, B.~Brooks, E.~Y.~Chew, and S.~M.~Meyers for their assistance with acquiring and analyzing the included datasets.  We also acknowledge V.~N.~Rajapakse and the anonymous referees for their valuable comments and suggestions that led to the improvement of this paper.

\ifCLASSOPTIONcaptionsoff
  \newpage
\fi



%

%


\bibliographystyle{IEEEtranS}
\bibliography{../biblio_ehler2}

\end{document}